\input{aipcheck}

\documentclass[final]{aipproc}

\layoutstyle{8x11single}

\usepackage{amsmath}
\usepackage{amssymb}
\usepackage{latexsym}
\usepackage{phonetic}

\newcommand{\be}{\begin{equation}}
\newcommand{\ee}{\end{equation}}
\newcommand{\beq}{\begin{equation}}
\newcommand{\eeq}{\end{equation}}
\newcommand{\bea}{\begin{eqnarray}}
\newcommand{\eea}{\end{eqnarray}}

\newcommand{\Tr}{\textrm{Tr}}

\newcommand{\xv}{{\mathbf x}}
\newcommand{\bra}{\langle}
\newcommand{\ket}{\rangle}
\newcommand{\re}{\textrm{Re}\,}
\newcommand{\im}{\textrm{Im}\,}
\newcommand{\eps}{\epsilon}
\newcommand{\id}{1\!\!\!1}
\newcommand{\dd}{\mbox{\hausad}}

\begin{document}

\title{QCD at nonzero chemical potential: recent progress \\ 
on the lattice\footnote{Based on a plenary talk at {\em XIth Quark Confinement and the Hadron Spectrum}, September 8-12, 2014, St.\ Petersburg, Russia.} 
}


\author{Gert Aarts}{
  address={Department of Physics, College of Science, Swansea University, Swansea SA2 8PP, United Kingdom}
}

\author{Felipe Attanasio}{
  address={Department of Physics, College of Science, Swansea University, Swansea SA2 8PP, United Kingdom},
  altaddress={CAPES Foundation, Ministry of Education of Brazil, Bras\'ilia - DF 70040-020, Brazil}
}

\author{Benjamin J\"ager}{
  address={Department of Physics, College of Science, Swansea University, Swansea SA2 8PP, United Kingdom}
}

\author{Erhard Seiler}{
  address={Max-Planck-Institut f\"ur Physik (Werner-Heisenberg-Institut), 80805 M\"unchen, Germany}
}

\author{D\'enes Sexty}{
  address={Department of Physics, University of Wuppertal, 42119 Wuppertal, Germany}
}

\author{Ion-Olimpiu Stamatescu}{
 address={Institut f\"ur Theoretische Physik, Universit\"at Heidelberg, 69120 Heidelberg, Germany}
}

\begin{abstract}
 We summarise recent progress in simulating QCD at nonzero baryon density using complex Langevin dynamics. After a brief outline of the main idea, we discuss gauge cooling as a means to control the evolution. Subsequently we present a status report for heavy dense QCD and its phase structure, full QCD with staggered quarks, and full QCD with Wilson quarks, both directly and using the hopping parameter expansion to all orders. 
 \end{abstract}

\maketitle

\section{Introduction}
\label{sec:intro}

\begin{figure}[b]
  \centerline{
    \includegraphics[width=0.6\textwidth]{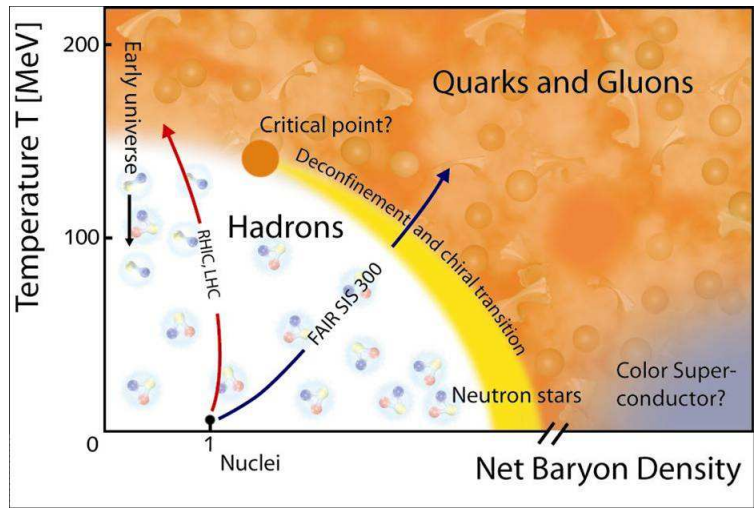} 
    }
   \caption{A possible sketch of the QCD phase diagram.
       } 
    \label{fig:pd}
\end{figure}

The QCD phase diagram, indicating the behaviour of strongly interacting matter as temperature and baryon density (or chemical potential) are varied, is under intense investigation, with the main motivation coming from the ongoing relativistic heavy-ion collisions at the Relativistic Heavy Ion Collider at Brookhaven and the Large Hadron Collider at CERN, the lower-energy studies at GSI/FAIR facilities, as well as from the need to understand compact astrophysical objects, such as neutron stars, from first principles. Furthermore, QCD describes one of the fundamental forces in Nature and hence there is an intrinsic desire to understand it under extreme conditions, such that the usual QCD vacuum is replaced by new phases of matter. A possible sketch of the QCD phase diagram is given in Fig.\ \ref{fig:pd}.

At this moment, a first-principles determination of the QCD phase diagram is still lacking. The reason is that in the regimes of interest, namely where the transitions to the quark-gluon plasma or to nuclear and quark matter take place, QCD is strongly coupled and hence a nonperturbative approach is needed. However, the applicability of the standard nonperturbative tool, lattice QCD, is severely hindered by the so-called sign problem, i.e.\ the fact that in the presence of a nonzero quark  (or baryon) chemical potential the fermion determinant is no longer real but complex,
\be
\label{eq1}
[\det M(\mu)]^* = \det M(-\mu^*).
\ee
This makes it hard, if not impossible, to assign probability weights to field configurations in numerical simulations of lattice QCD.

A number of approaches to evade the sign problem is currently under investigation. These will not be reviewed here, see instead e.g.\ Refs. \cite{deForcrand:2010ys,Aarts:2013naa,Gattringer:2014nxa}. Instead we focus on complex Langevin dynamics, which has recently for the first time been applied to full QCD \cite{Sexty:2013ica,Aarts:2014bwa}. 
Some alternative reviews can be found in Refs.\ \cite{Aarts:2013bla,Aarts:2013uxa,Sexty:2014dxa}.

\section{Complex Langevin dynamics}
\label{sec:cl}

\begin{figure}[b]
  \centerline{
    \includegraphics[width=0.4\textwidth]{plot-rho.eps} $\;\;\;\;$
    \includegraphics[width=0.4\textwidth]{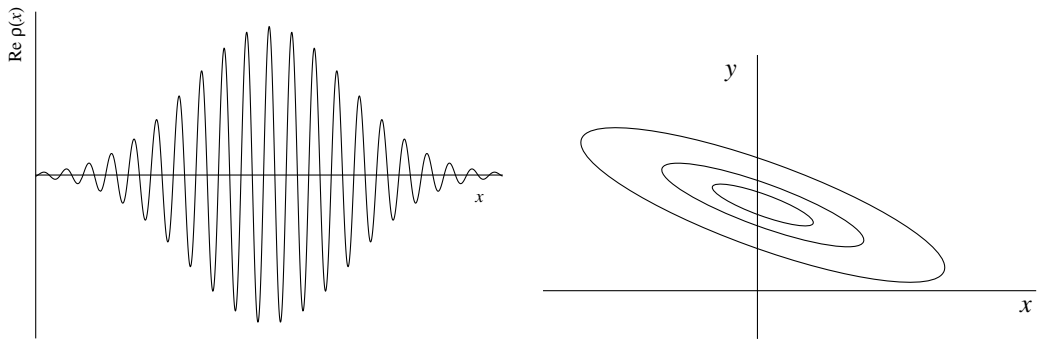}
    }
   \caption{A complex, highly oscillating distribution $\rho(x)$ (left) might be represented by a real and nonnegative distribution $P(x,y)$ in the complexified space (right).   
    } 
    \label{fig:dist}
\end{figure}

Consider the QCD partition function, 
\be
Z = \int DU D\bar\psi D\psi\, e^{-S} = \int DU\, e^{-S_{\rm YM}}\det M(\mu),
\ee
where in the final expression the quark fields have been integrated out, resulting in the complex determinant. We denote the weight under the integral generically as $\rho(x)$, where $x$ indicates all remaining field dependence. At nonzero chemical potential, this weight is complex and highly oscillating, see Fig.\ \ref{fig:dist} (left). Hence it is not obvious what the dominant configurations in the path integral are. Simply ignoring the complexity, and using e.g.\ the phase-quenched weight $|\rho(x)|$, will lead to a severe overlap problem, since the full and phase-quenched ensembles describe manifestly different physical theories.

The main idea underlying complex Langevin dynamics is that there exists a real and nonnegative weight $P(x,y)$ in the complex plane, or complexified configuration space, such that
\be
\label{eq:rho}
\int dx\,\rho(x)O(x) = \int dx dy\, P(x,y)O(x+iy),
\ee
for holomorphic observables $O(x)$, see Fig.\ \ref{fig:dist} (right). The question is how such a probability weight can be constructed.

In complex Langevin dynamics \cite{parisi,klauder}
 $P(x,y)$ is effectively obtained as the solution of a stochastic process, which takes place in the complex plane. 
  For one degree of freedom, with partition function
  \be
  Z = \int dx\, e^{-S(x)}, \quad\quad\quad S(x)\in\mathbb{C}, 
  \ee
  the complex Langevin equations take the form
  \be
  \partial_t x = -\re \partial_zS(z) +\eta, 
  \quad\quad\quad
  \partial_t y = -\im \partial_zS(z),
  \ee
  where $S(z)=S(x+iy)$, $t$ is the Langevin time,  and the noise satisfies
\be
\bra\eta(t)\ket=0,
  \quad\quad\quad
\bra\eta(t)\eta(t')\ket=2\delta(t-t').
\ee
The proof of the applicability of this method \cite{Aarts:2009uq,Aarts:2011ax}
 goes via the Fokker-Planck equation for the associated distribution,
\be
\label{eq:FPE}
\partial_t P(x,y;t) = \left[\partial_x(\partial_x+\re\partial_zS)+\partial_y\im\partial_zS\right] P(x,y;t),
\ee
and the conjectured relation (\ref{eq:rho}). In the case of holomorphic actions, the method is reliable, provided that the equilibrium distribution $P(x,y)$ is well localised in the imaginary direction and certain criteria for correctness, which can be verified a posteriori, are satisfied \cite{Aarts:2009uq,Aarts:2011ax}.
For meromorphic drifts, i.e.\  drifts $-\partial_zS$ with poles, problems may appear but not necessarily so \cite{Mollgaard:2013qra,Greensite:2014cxa}.
We note that no importance sampling is needed, since equilibrium is reached as in Brownian motion.
An explicit solution of Eq.\ (\ref{eq:FPE}) is unfortunately only available in a few selected cases (see e.g.\ Refs. \cite{Nakazato:1985zj,Haymaker:1989hn,Duncan:2013wm,Aarts:2013uza}) and hence the analysis of $P(x,y)$ requires its explicit construction using the Langevin equation.

\section{Gauge theories}
\label{sec:gt}

\begin{figure}[t]
  \centerline{
    \includegraphics[width=0.5\textwidth]{plot_fsu3_truudagger_Nt4_ch2_1outof4.eps} 
    \includegraphics[width=0.45\textwidth]{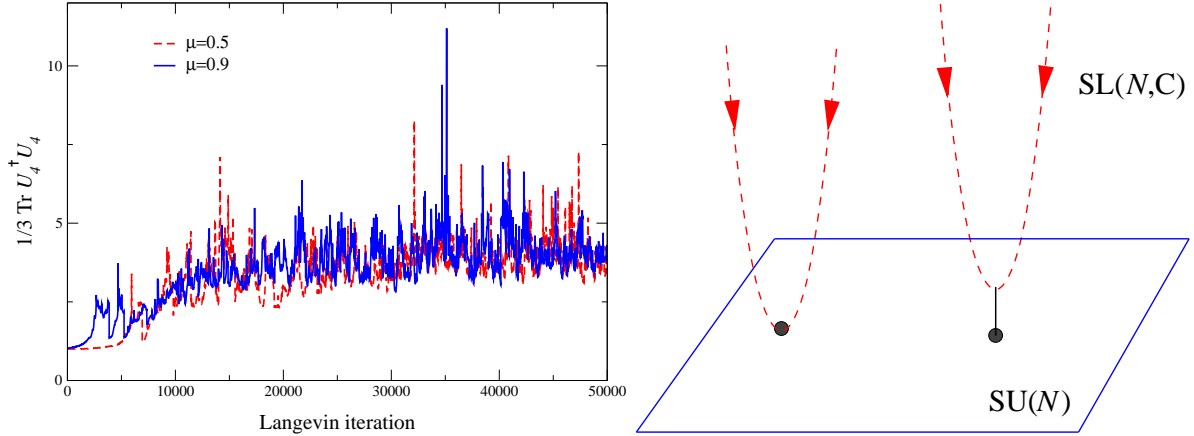} 
        }
   \caption{Left: Langevin evolution of a unitarity norm, $\Tr\, U_4U_4^\dagger/3$, in heavy dense QCD on a $4^4$ lattice at $\beta=5.6$, $\kappa=0.12$ and $N_f=2$ for two values of the chemical potential $\mu$,  without gauge cooling \cite{Aarts:2008rr}.
Right: Gauge cooling reduces the unitarity norm and brings a configuration in SL($N,\mathbb{C}$) as close as possible to SU($N$). 
       } 
    \label{fig:truu}
\end{figure}

For nonabelian SU($N$) gauge theories on the lattice, the Langevin update can be written for gauge links $U$ (suppressing most indices) as \cite{batrouni,Aarts:2008rr}
\be
 U(t+\eps) = R(t)U(t),
 \quad\quad\quad\quad
 R = \exp\left[ -i\sum_a\lambda_a\left( \eps D_aS +\sqrt{\eps}\eta_a\right)\right],
\ee
where $\eps$ is the Langevin stepsize and $\lambda_a$ are the Gell-Mann matrices ($a=1,\ldots,N^2-1$).
For QCD the action includes the logarithm of the determinant, and hence the drift has a pole where the determinant vanishes. Whether this leads to problems in practice is still under investigation; for the results shown below, we believe this is not the case.
All matrices above have determinant 1, however, when the action and hence the drift are complex, they are no longer unitary. Complex Langevin dynamics takes therefore place in SL($N,\mathbb{C}$), the complex extension of SU($N$).
This can be demonstrated by considering unitarity norms, i.e.\ norms which measure the distance from SU($N$). Examples of these are \cite{Aarts:2013uxa}
\be
\dd_1 = \frac{1}{N}\Tr\left( UU^\dagger -\id\right)
\quad\quad\quad\quad
\dd_2 = \frac{1}{N}\Tr\left( UU^\dagger -\id\right)^2,
\ee
etc. For unitary matrices, these norms vanish, while for nonunitary SL($N,\mathbb{C}$) matrices, they exceed 0. An example is given in Fig.\ \ref{fig:truu} (left) where the evolution of the unitarity norm $\Tr\, U_4^\dagger U_4/3\geq 1$ is shown in the case of heavy dense QCD (HDQCD), to be discussed further below. This figure is taken from Ref.\ \cite{Aarts:2008rr}, in which complex Langevin dynamics was first applied to HDQCD.

In order to satisfy the criteria alluded to above and make sure the Langevin process explores the enlarged configuration space in a controlled fashion, it is necessary to restrain the distance from SU($N$). This can be achieved with gauge cooling, in which Langevin updates are interspersed with gauge transformations that reduce the unitarity norms but do not affect observables \cite{Seiler:2012wz}. Gauge cooling affects the Langevin process as a whole, since the Langevin update and the cooling step do not commute. 
For a link $U_{x,\mu}$, gauge transformations in SL($N,\mathbb{C}$) take the form
\be
U_{x,\mu} \rightarrow \Omega_xU_{x,\mu} \Omega_{x+\hat\mu}^{-1}, 
\quad\quad\quad\quad
\Omega_x = e^{i\omega_{ax}\lambda_a},
\quad\quad\quad\quad
\omega_{ax}\in\mathbb{C}.
\ee
By choosing $\omega_{ax}=i\alpha f_{ax}$ purely imaginary, gauge cooling does not affect the unitary subgroup but only the distance in the orthogonal direction. Linearising in $\alpha>0$ indeed shows that e.g.\ the norm $\dd_1$ is reduced: after a gauge transformation at site $x$,  $\dd_1'-\dd_1 = -(\alpha/N) f_{ax}^2 <0$. If a configuration is gauge-equivalent to an SU($N$) configuration, gauge cooling will return it to the unitary group. If not, there is a minimal distance. This is illustrated in Fig.\ \ref{fig:truu} (right).
Gauge cooling can be implemented adaptively \cite{Aarts:2013uxa}, which allows for a significant speed-up of the simulations \cite{Aarts:2014kja}.
We emphasise that in QCD the unitary submanifold is very unstable and hence gauge cooling is essential. Besides this, it is also necessary to use an adaptive stepsize during the Langevin update \cite{Aarts:2009dg}.

\section{Heavy dense QCD}
\label{sec:hdm}

As a first application we consider QCD with static quarks at nonzero chemical potential, known as heavy dense QCD (HDQCD)
\cite{Bender:1992gn,Blum:1995cb}. Starting from Wilson quarks with hopping parameter $\kappa$, it can be obtained by discarding all spatial hopping terms, keeping only temporal hopping.
In this case the fermion determinant takes the form
\be
\det M = \prod_\xv \det\left(1+h e^{\mu/T}{\cal P}_\xv\right)^2 \det\left(1+h e^{-\mu/T}{\cal P}_\xv^{-1}\right)^2,
\ee
where $h=(2\kappa)^{N_\tau}$, ${\cal P}_\xv$ and ${\cal P}_\xv^{-1}$ are the (conjugate) Polyakov loops and the remaining determinants are in colour-space only. 
For the gauge fields, the full Wilson gauge action, with coupling $\beta$, is included. Hence this approximation goes beyond e.g.\ the strong coupling expansion considered in Refs.\ \cite{Fromm:2011qi,Fromm:2012eb,Langelage:2014vpa}.
The determinant still satisfies Eq.\ (\ref{eq1}).

\begin{figure}
 \centerline{
    \includegraphics[width=0.48\textwidth]{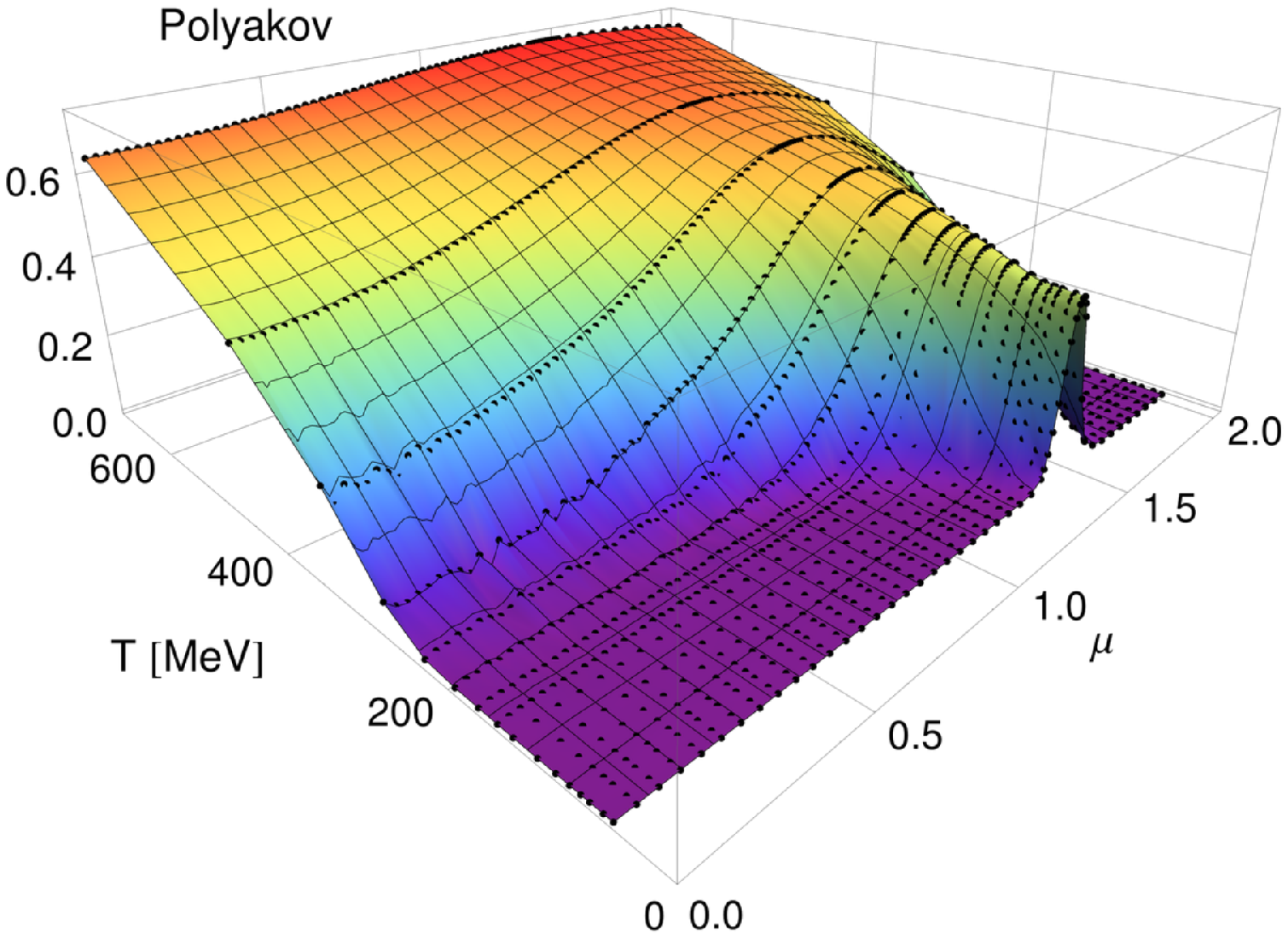} 
    \includegraphics[width=0.48\textwidth]{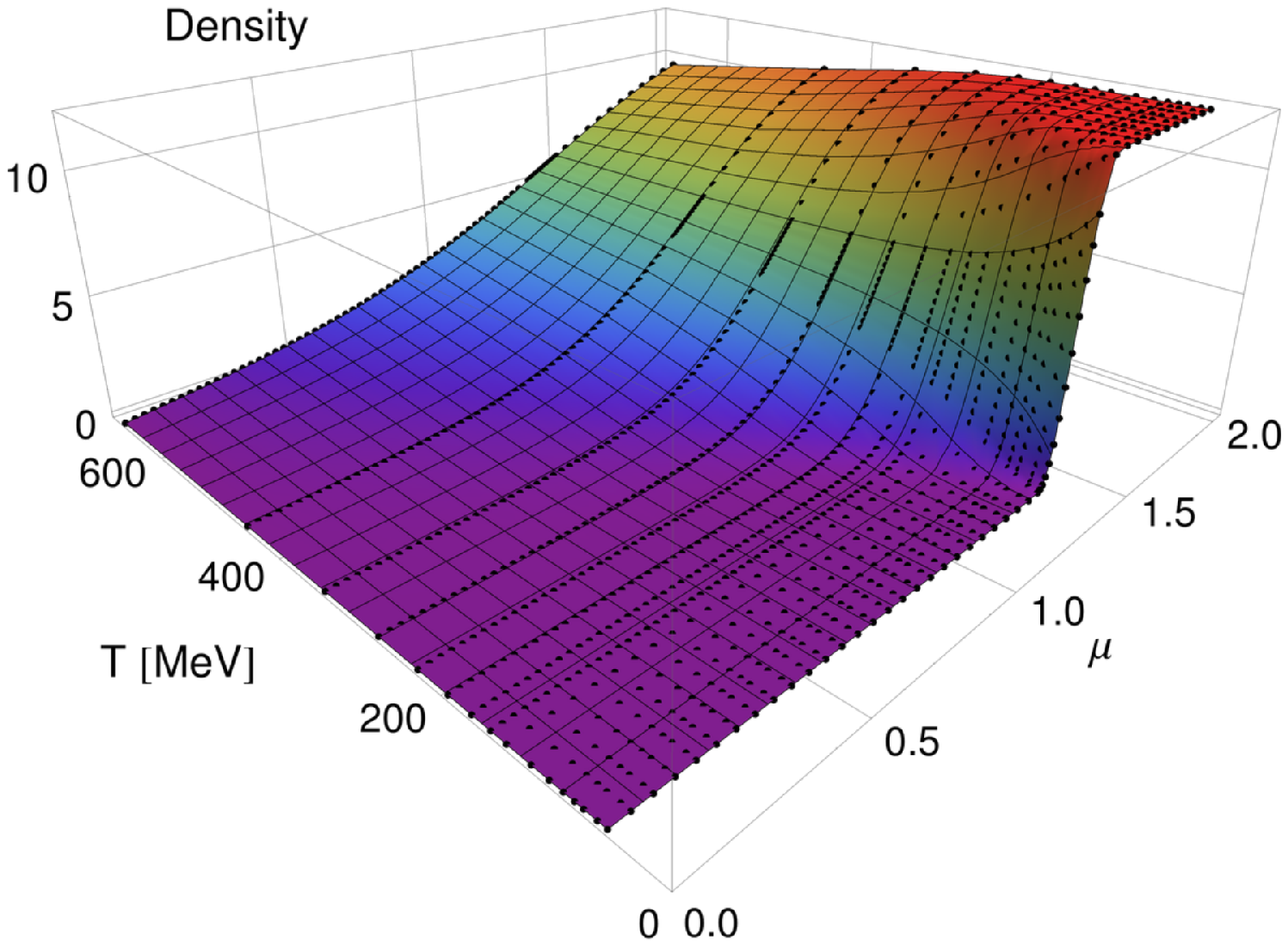} 
   }
  \caption{HDQCD: Polyakov loop (left) and density (right) in the plane of temperature (in MeV) and chemical potential (in lattice units) on a $8^3\times N_\tau$ lattice at $\beta=5.8$, $\kappa=0.12$ and $N_f=2$.
          } 
    \label{fig:hd1}
\end{figure}

This theory has a nontrivial phase diagram: there is a thermal deconfinement transition as the temperature is increased (by varying $N_\tau$ at fixed $\beta$ or vice versa), just as in the pure glue theory. This transition is first order for very heavy quarks (small $\kappa$) and a crossover for slightly lighter but still heavy quarks. As the chemical potential is increased at fixed temperature, there is a transition to the high-density state. At zero temperature, this transition is expected to be first order and to occur at $h e^{\mu/T}=1$, or $\mu=\mu_c\equiv m_q$, where in this model the quark mass is given by $m_q\equiv -\ln(2\kappa)$. This is clearly a simplification with respect to full QCD and can be improved by including higher-order terms in $\kappa$, see e.g.\ Refs.\ \cite{Aarts:2002,DePietri:2007ak,Langelage:2014vpa,Aarts:2014bwa} and below. The two transitions are expected to meet in the $T-\mu$ plane.

\begin{figure}
 \centerline{
    \includegraphics[width=0.48\textwidth]{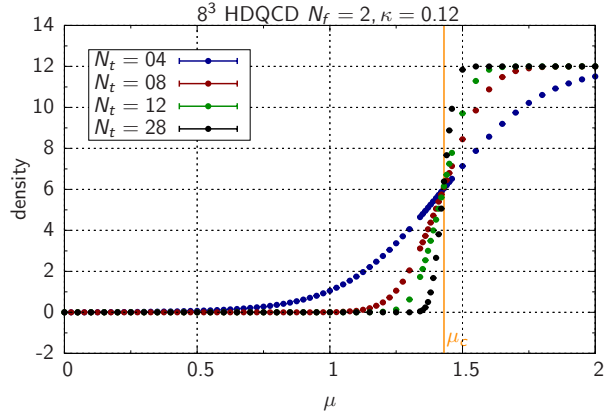} 
   }
  \caption{Silver Blaze feature in HDQCD: behaviour of the density with decreasing temperature (increasing $N_\tau$). The critical chemical potential is $\mu_c=- \ln(2\kappa) =1.43$ and the saturation density is $n_{\rm sat}=12$. Other parameters as in Fig.\ \ref{fig:hd1}.
          } 
    \label{fig:hd2}
\end{figure}
\begin{figure}
 \centerline{
    \includegraphics[width=0.8\textwidth]{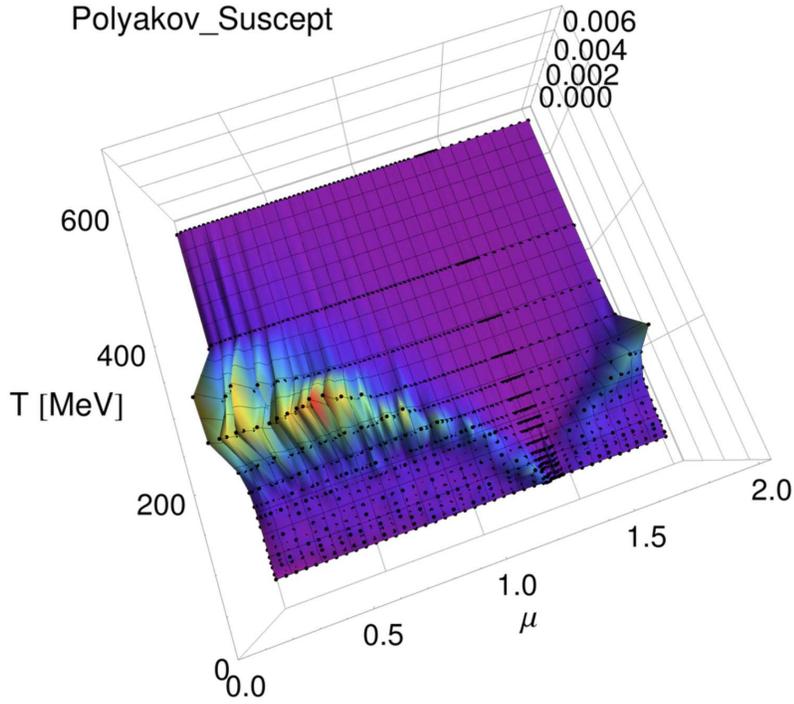} 
   }
  \caption{HDQCD: Polyakov loop susceptibility in the $T-\mu$ plane. Parameters as in Fig.\ \ref{fig:hd1}.
          } 
    \label{fig:hd3}
\end{figure}

Here we present preliminary results for $N_f=2$ flavours at fixed $\beta=5.8$ (corresponding to $a\sim 0.15$ fm; we used the gradient flow to set the scale, see e.g.\ Ref.\ \cite{Borsanyi:2012zs}) and $\kappa=0.12$ (corresponding to $\mu_c\sim 1.43$) on lattices of size $8^3\times N_\tau$, with $N_\tau=2-28$ (see also Ref.\ \cite{Aarts:2014kja} for results at $\kappa=0.04$).
 The results for the Polyakov loop and the density are shown in Fig.\ \ref{fig:hd1}. The Polyakov loop is close to zero in the low $T$-low $\mu$ corner of the $T-\mu$ plane and then increases as the temperature and/or chemical potential is increased, signalling deconfinement. 
 The density rises as the chemical potential is increased and the rise is steeper at low temperature. This is an indication of the expected behaviour in the Silver Blaze region \cite{Cohen:2003kd}: 
  at low temperature the density rises sharply, jumping discontinuously as the temperature is taken to zero, see Figs.\  \ref{fig:hd1} (right) and \ref{fig:hd2}. The maximum value the density can take is saturation density, $n_{\rm sat}=2N_fN_c=12$, for which all sites on the lattice are maximally occupied. We note that in the static limit there is no distinction  between $m_\pi/2$ and $m_B/3$, and hence the Silver Blaze problem is easier than in the full QCD.
  
 At the critical chemical potential and at low temperature, the Polyakov loop has a maximum after which it drops again to zero. This is a lattice artefact arising from the saturation of the density. When all sites on the lattice are occupied, the Polyakov loop becomes again insensitive to the chemical potential (inverse Silver Blaze problem) and drops to zero. Hence the region where $\mu>\mu_c$ is unphysical.

The Polyakov loop susceptibility is shown in Fig.\ \ref{fig:hd3}.  From its behaviour one can clearly see the emergence of the phase boundary in the $T-\mu$ plane. In order to make this more precise, simulations at larger spatial volume are required, which are currently in process. Finally, we observe that the region where $\mu>\mu_c$ is ``dual'' to the low-$\mu$ region, in the sense that the roles of empty and filled states is reversed.

\section{Full QCD}
\label{sec:full}

Last year first results for full QCD have finally appeared \cite{Sexty:2013ica}. The fermion determinant leads to an additional term in the drift, which requires the inverse of the fermion matrix (though not its determinant). The inversion is done stochastically, using conjugate gradient, and is the most expensive part of the computation. Gauge cooling and adaptive stepsize are needed. So far the approach has been implemented for unimproved staggered fermions \cite{Sexty:2013ica} and standard Wilson fermions \cite{Aarts:2014bwa}. During the simulations, the unitarity norms, distributions and eigenvalues of the Dirac operator near zero are being monitored. In order to have trust in the results, comparisons with HDQCD and reweighting are done, where possible. A comparison with the hopping parameter expansion to all orders is discussed below.

\begin{figure}[b]
 \centerline{
    \includegraphics[width=0.49\textwidth]{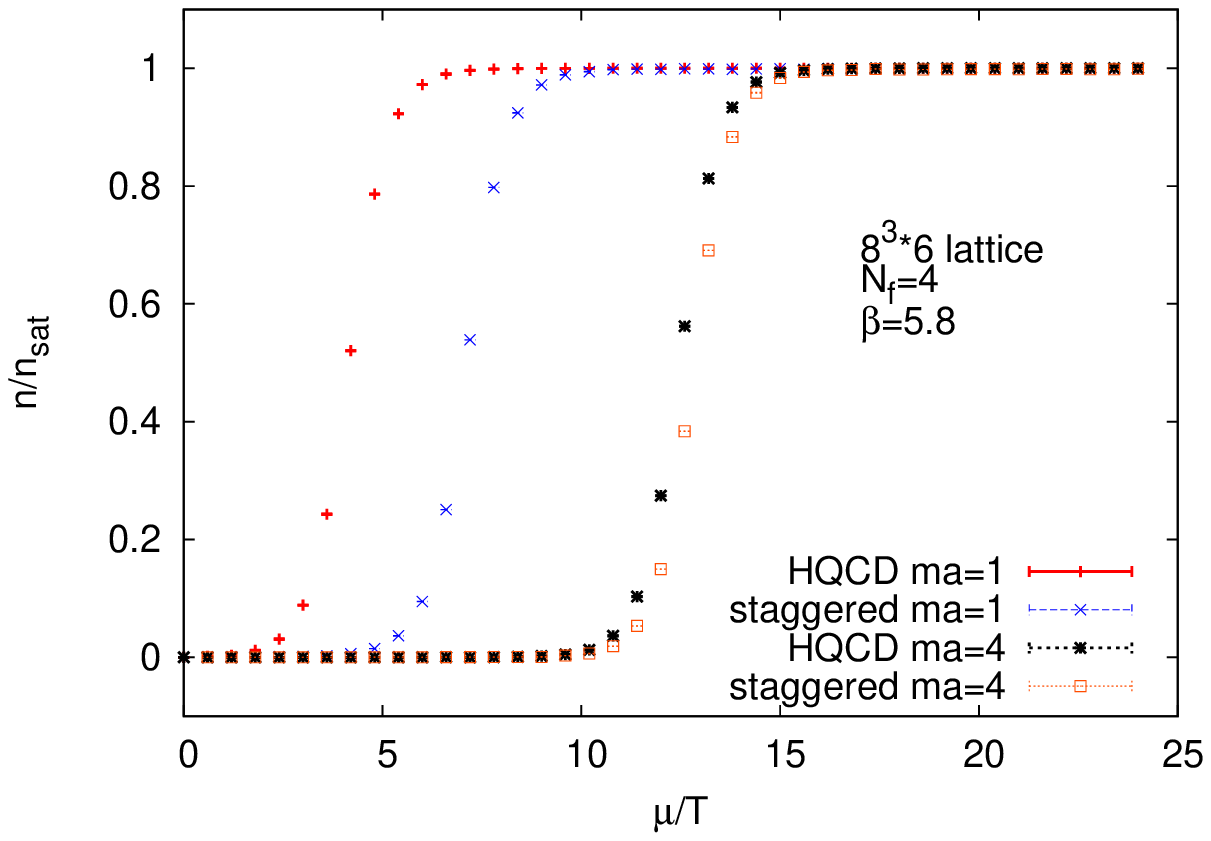} 
     \includegraphics[width=0.49\textwidth]{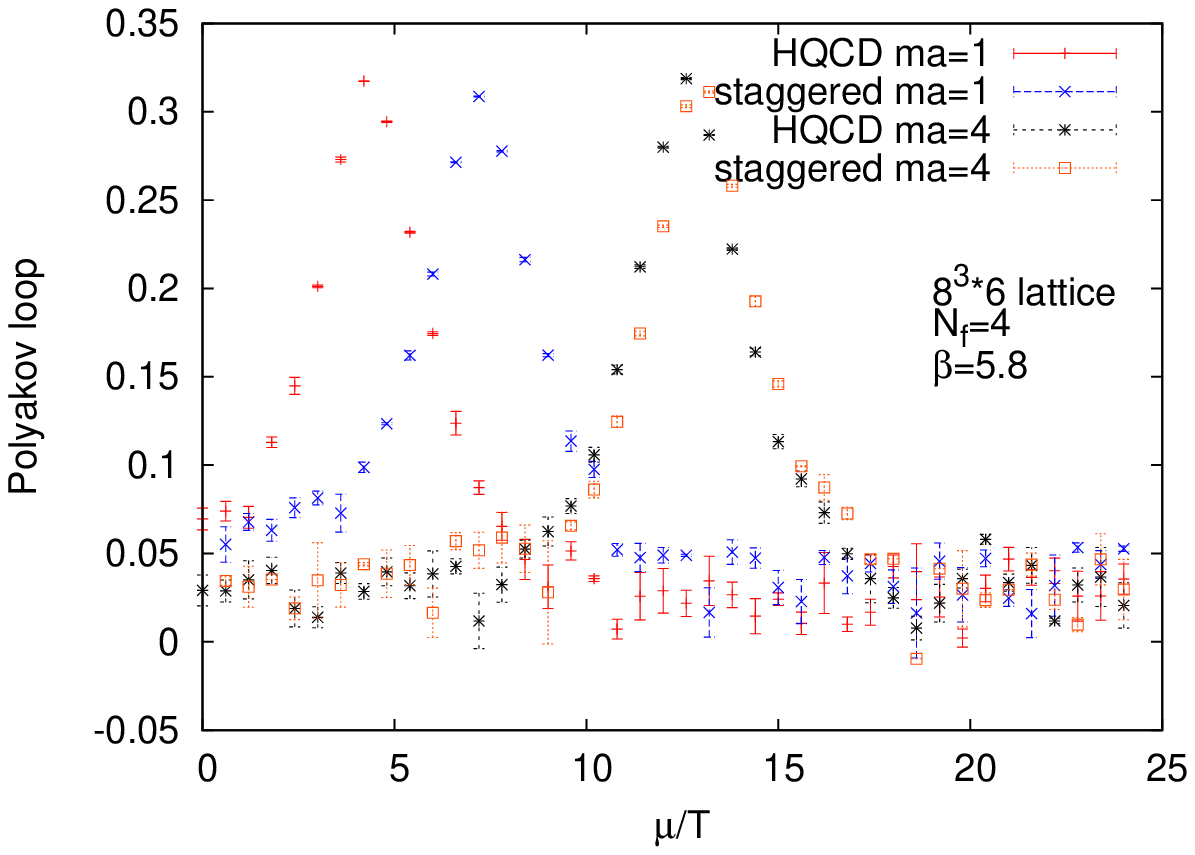} 
   }
  \caption{Density $n/n_{\rm sat}$ (left) and Polyakov loop (right) as a function of $\mu/T$ in QCD with $N_f=4$ flavours of staggered quarks on a $8^3\times 6$ lattice at $\beta=5.8$, for two values of the bare quark mass, $am=1$ and $4$   \cite{Sexty:2013ica}. The result in HDQCD is also shown. 
            } 
    \label{fig:full}
\end{figure}

In Fig.\ \ref{fig:full} both the density, normalised with the saturation density, and the Polyakov loop are shown, for four flavours of unimproved staggered fermions using two bare quark masses, $am=1$ and $am=4$, on a $8^3\times 6$ lattice. Also shown are the results from  the static limit, as in HDQCD, using a $1/m$ expansion. The latter can be compared with the $am=4$ data, which corresponds to heavy quarks as well. For the density we observe the rise from zero to saturation as the chemical potential is increased. For the large mass value, there is agreement between the static limit and the full theory, indicating that the quarks are indeed very heavy and providing an important crosscheck for both approaches.  For the lighter staggered quarks, there is a clear deviation which is due to the lack of dynamics in the static limit. The Polyakov loop shows qualitatively the same behaviour as above: i.e. when saturation is reached, its value drops down close to zero. Hence only the region before the Polyakov loop reaches its maximum (i.e.\ before half-filling) is physical. 

To summarise, it is satisfying to see agreement between the full theory and the static limit for the large mass parameter and disagreement for the lighter quarks. The behaviour of the density and the Polyakov loop is as expected. The important open question concerns details of the onset at low temperature, which is currently under investigation.

\section{Hopping parameter expansion}
\label{sec:hop}

In order to bridge the gap between the static limit and the full theory, we have recently proposed the hopping parameter expansion to all orders  \cite{Aarts:2014bwa}. The static limit clearly has some shortcomings, such as  immediate saturation after onset at $T=0$ and coincidence of  $m_B/3$ and $m_\pi/2$. These limitations are already overcome by including the lowest ${\cal O}(\kappa^2)$ \cite{Aarts:2002,DePietri:2007ak} and ${\cal O}(\kappa^4)$ \cite{Fromm:2011qi,Fromm:2012eb,Langelage:2014vpa,Langelage:2013paa} corrections. However, an extension to higher order is typically quite involved. Here we discuss two expansions which are systematic and can be truncated at high order; we consider contributions up to ${\cal O}(\kappa^{50})$ below. Since we start from the same fermion and gauge action, a comparison with the full theory is straightforward, namely simply by choosing the same bare parameters. 
The determinants in the hopping parameter expansion are still complex and hence we use complex Langevin dynamics to solve the truncated theory as well. A comparison provides therefore again an important crosscheck between the different approaches.

We consider two expansions: one in the hopping parameter $\kappa$ for both spatial and temporal terms, and one in the spatial hopping parameter $\kappa_s$ only. The starting point for the latter is HDQCD: hence all the chemical potential dependence is already included at lowest order and the role of the higher order terms is to contribute more kinetic terms. The former approach is numerically cheaper but includes terms of the order ${\cal O}(\kappa^n e^{n\mu})$ at the $n$th term, affecting the expansion at larger values of $\mu$. 

To wit, in the straightforward hopping expansion we write
\be
\det M  = \det\left(1-\kappa Q\right) = \exp \sum_{n=1}^\infty -\frac{\kappa^n}{n}\Tr\, Q^n,
\ee
whereas in the $\kappa_s$ expansion the heavy dense determinant is factored out first,
\be
\det M  = \det\left(1-R-\kappa_s S\right) =\det\left(1-R\right)  \exp \sum_{n=1}^\infty -\frac{\kappa_s^n}{n}\Tr \left(\frac{1}{1-R}S\right)^n,
\ee
where we decomposed the fermion matrix as $M=1-\kappa Q = 1-\kappa_s S-R$, with $S/R$ containing hoppings in the spatial/temporal direction only. The $\kappa_s$ expansion requires the inversion of the heavy dense matrix, which can be achieved analytically, while the $\kappa$ expansion requires no inversion at all. The traces appearing in the Langevin drift are estimated stochastically.

\begin{figure}[b]
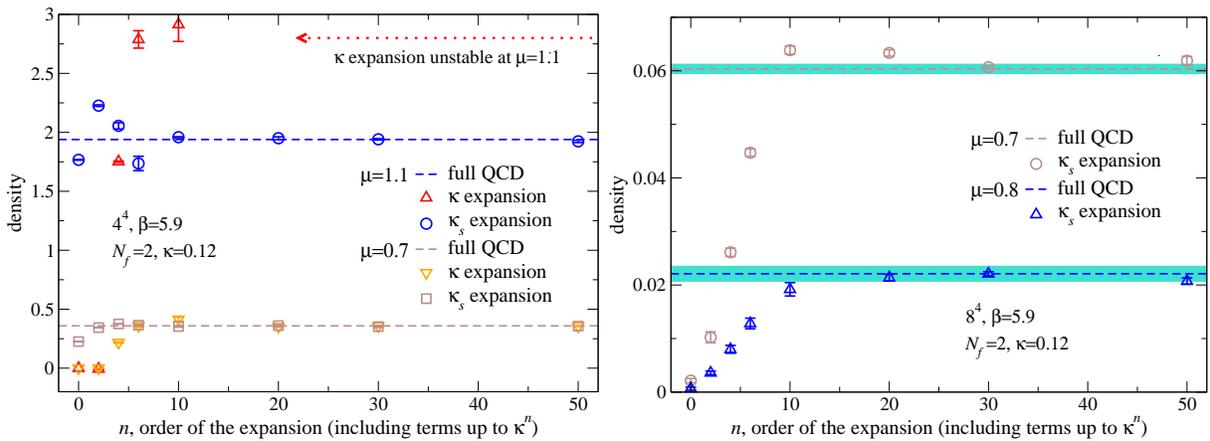

 \centerline{
    \includegraphics[width=0.48\textwidth]{plot-dens-4x4-kappa0.12.eps} 
     \includegraphics[width=0.49\textwidth]{plot-dens-8x8-kappa0.12-mu78.eps} 
   }
  \caption{Dependence of the density (in lattice units) on the order of the truncation in the hopping expansion on a $4^4$ lattice (left) and an $8^4$ lattice (right) in QCD with $N_f=2$ flavours of Wilson quarks, at $\beta=5.9$ and $\kappa=0.12$, for various values of the chemical potential \cite{Aarts:2014bwa}. The result from the $\kappa_s$-expansion is always shown, while the $\kappa$-expansion may break down at larger $\mu$ values and lower temperatures. Also shown are the results from direct complex Langevin simulations of full QCD.
            } 
    \label{fig:hopp}
\end{figure}

Some results are shown in Fig.\ \ref{fig:hopp}. In each case we show the density for two values of $\mu$ as a function of the order of the expansion, up to order 50. Also shown are the results obtained with complex Langevin simulations of the full theory with Wilson fermions. The left figure has results on a $4^4$ lattice, for both the $\kappa_s$ and the $\kappa$ expansion. The latter breaks down for the larger $\mu$ value. The right figure is for an $8^4$ lattice, for the $\kappa_s$ expansion only. In each case we find excellent agreement with the full result, thereby providing justification for both methods. We also observe convergence of the $\kappa_s$ expansion, to the correct result. 
In view of potential problems associated with meromorphic drifts, we note here that in the $\kappa_s$ expansion poles still exist, albeit of higher order, while in the $\kappa$ expansion poles are completely absent. Hence the agreement between the various approaches indicates,  for the parameter values chosen here, that problems coming from poles associated with the $\log\det$ in the action are not present.

In this case the remaining questions concern the convergence of the expansions for lighter quarks (larger $\kappa$) and at lower temperatures. Again this is currently under investigation.

\section{Summary}
\label{sec:out}

In this contribution we have given a brief overview of the status of the applicability of complex Langevin dynamics to QCD at nonzero chemical potential. There is progress on a number of fronts, most importantly the possibility to carry out simulations for full QCD with staggered and Wilson fermions. This has opened up the door towards addressing questions with regard to the QCD phase diagram. In order to test the approach, we have given first results for QCD in the heavy dense limit, where the phase diagram indeed can be determined by direct simulation. For full QCD, the important open questions relate to the feasibility of simulating light quarks at low temperature and the role of poles in the drift arising from the logarithm of the determinant in the action. These questions can partly be answered by comparison with the hopping parameter expansion to all orders.

\begin{theacknowledgments}
 GA thanks the organisers for an excellent meeting, and Chris Allton and Peter Arnold for discussion.
GA is supported by STFC, the Royal Society, the Wolfson Foundation and the Leverhulme Trust.
FA acknowledges financial support from CAPES Foundation via the Science without Borders programme, scholarship No. Bex 9463/13-5.
BJ is supported by STFC.
ES and IOS are supported by the Deutsche Forschungsgemeinschaft.  
For computational support we thank  BMBF and MWFK Baden-W\"urttemberg (bwGRiD cluster), HPC Wales, and the STFC funded DiRAC Facility. 
\end{theacknowledgments}

\bibliographystyle{aipproc}

\end{document}